\documentclass[sigconf,natbib=true]{acmart}

\usepackage{enumitem}
\usepackage{amsmath}
\usepackage{mathtools}
\usepackage{multirow}
\newtheorem{prop}{Proposition}
\DeclareMathOperator*{\argmax}{arg\,max}
\DeclareMathOperator*{\argmin}{arg\,min}
\usepackage[linesnumbered,ruled]{algorithm2e}
\AtBeginDocument{%
  \providecommand\BibTeX{{%
    \normalfont B\kern-0.5em{\scshape i\kern-0.25em b}\kern-0.8em\TeX}}}

\copyrightyear{2023} 
\acmYear{2023} 
\setcopyright{acmlicensed}\acmConference[RecSys '23]{Seventeenth ACM Conference on Recommender Systems}{September 18--22, 2023}{Singapore, Singapore}
\acmBooktitle{Seventeenth ACM Conference on Recommender Systems (RecSys '23), September 18--22, 2023, Singapore, Singapore}
\acmPrice{15.00}
\acmDOI{10.1145/3604915.3608771}
\acmISBN{979-8-4007-0241-9/23/09}

\begin{document}

% \title{SharpCF: Simple Yet Effective Adversarial Training for Collaborative Filtering 
% }

\title{ Adversarial  Collaborative Filtering for Free 
}

\author{Huiyuan Chen}
\email{hchen@visa.com}
\affiliation{%
  \institution{Visa Research}
  \city{Palo Alto}
  \state{CA}
  \country{USA}
}

\author{Xiaoting Li}
\email{xiaotili@visa.com}
\affiliation{
  \institution{Visa Research}
  \city{Palo Alto}
  \state{CA}
  \country{USA}
}

\author{Vivian Lai}
\email{viv.lai@visa.com}
\affiliation{
  \institution{Visa Research}
  \city{Palo Alto}
  \state{CA}
  \country{USA}
}

\author{Chin-Chia Michael Yeh}
\email{miyeh@visa.com}
\affiliation{%
  \institution{Visa Research}
  \city{Palo Alto}
  \state{CA}
  \country{USA}
}

\author{Yujie Fan}
\author{Yan Zheng}
\email{yazheng@visa.com}
\affiliation{%
  \institution{Visa Research}
  \city{Palo Alto}
  \state{CA}
  \country{USA}
}

% \author{Yan Zheng}
% \email{yazheng@visa.com}
% \affiliation{%
%   \institution{Visa Research}
%   \city{Palo Alto}
%   \state{CA}

\author{Mahashweta Das}
\author{Hao Yang}
 \email{haoyang@visa.com}
\affiliation{
  \institution{Visa Research}
  \city{Palo Alto}
  \state{CA}
  \country{USA}
}

% \author{Hao Yang}
%  \email{haoyang@visa.com}
% \affiliation{
%   \institution{Visa Research}
%   \city{Palo Alto}
%   \state{CA}

\renewcommand{\shortauthors}{Huiyuan Chen, et al.}

\begin{abstract}
Collaborative Filtering (CF) has been successfully used to help users discover the items of interest. Nevertheless, existing CF methods suffer from noisy data issue, which negatively impacts the quality of recommendation. To tackle this problem, many  prior studies leverage  adversarial learning  to regularize the representations of users/items, which  improves both generalizability and robustness. Those methods often learn adversarial perturbations and model parameters under min-max optimization framework. However, there still have two major drawbacks: 1) Existing methods lack theoretical guarantees of why adding perturbations improve the model generalizability and robustness; 2)  Solving min-max optimization is time-consuming.  In addition to updating the model parameters, each iteration requires additional computations to update the perturbations, making them not scalable for industry-scale datasets.

In this paper, we present Sharpness-aware Collaborative Filtering  (SharpCF), a simple yet effective method that conducts adversarial training without extra computational cost over the base optimizer. To achieve this goal, we first revisit the existing adversarial collaborative filtering and discuss its connection with recent Sharpness-aware Minimization. This analysis shows that adversarial training actually seeks model parameters that lie in neighborhoods around the optimal model parameters having uniformly low loss values, resulting in better generalizability. To reduce the computational overhead, SharpCF introduces a novel trajectory loss to measure the alignment between current weights and past weights. Experimental results on real-world datasets demonstrate that our SharpCF achieves superior performance with almost zero additional computational cost comparing to adversarial training. 
\end{abstract}



\maketitle

\section{Introduction}

Recommender systems have become increasingly popular due to their ability to filter information and provide personalized recommendations to users on the web~\cite{covington2016deep, koren2009matrix, rendle2009bpr}. One of the most prominent techniques used in recommender systems is Collaborative Filtering (CF)\cite{he2017neural, koren2008factorization}. CF considers users' historical interactions and assumes that users who have shared similar preferences in the past tend to make similar decisions in the near future. To achieve this goal, most CF methods learn an encoder to embed users and items into a shared space and then optimize an objective function to learn informative user and item representations. CF-based methods have been successfully deployed in industries due to their simplicity and effectiveness, such as Factorization Machines\cite{guo2017deepfm,lian2018xdeepfm} and Graph Neural Networks~\cite{ying2018graph,he2020lightgcn}.

However, many state-of-the-art collaborative filtering (CF) methods remain vulnerable to noisy data, and their performance can degrade significantly under unnoticeable perturbations~\cite{wang2021clicks,tian2022learning,chen2022denoising,chen2023sharp}. Indeed, real-world data is often noisy, where user preferences may not necessarily align with the interacted items. For instance, a significant portion of purchases may result in negative reviews or returns. Such false-positive interactions hinder a model from learning the actual user preferences, leading to low-quality recommendations~\cite{wen2019leveraging,wang2021denoising}. To tackle this problem, many prior studies leverage the adversarial learning principle to regularize the representations of users and items~\cite{he2018adversarial,yuan2019adversarial,wu2021fight,du2022understanding,suresh2021adversarial}. These methods employ a min-max optimization framework to alternatively learn adversarial perturbations and model parameters. For instance, in the APR model~\cite{he2018adversarial}, adversarial training is applied to a Matrix Factorization model by directly adding adversarial perturbations to the embedding vectors of both users and items.

While existing adversarial collaborative filtering methods have shown promising results, there still have two major limitations: 1) These methods lack theoretical guarantees on why the addition of adversarial perturbations improves the model's generalizability and robustness. Interpretability remains unclear as noisy data is inherently different from adversarial perturbations, which are carefully crafted modifications to the original data~\cite{ian}; 2) The process of solving min-max optimization is time-consuming~\cite{Wong2020Fast}. In each iteration, besides updating the model parameters (e.g., the outer minimization problem), additional computations are required to update the learnable  perturbations (e.g., the inner maximization problem). This makes the existing approaches unsuitable for learning with large-scale datasets. Therefore, developing more efficient and interpretable adversarial collaborative filtering methods remains a significant challenge in  recommender systems.

\textbf{Present Work.} To overcome the limitations outlined above,  we propose Sharpness-aware  Collaborative Filtering (SharpCF), a simple yet effective method that enables adversarial training without incurring extra computational costs over the base optimizer. To achieve this goal, we first revisit existing adversarial collaborative filtering techniques and discuss their connection with recent developed Sharpness-aware Minimization~\cite{foret2021sharpnessaware,kwon2021asam,andriushchenko2022towards,du2022sharpness}. Our analysis reveals that adversarial training actually seeks model parameters that lie in neighborhoods around the optimal model parameters with uniformly low loss values. In other words, the adversarial training favors \textsl{flat} minima rather than \textsl{sharp} minima, resulting in better model generalizability. 

To reduce the computational overhead, our SharpCF introduces a novel trajectory loss that measures the alignment between current  and past model states. We further  demonstrate that this trajectory loss can prevent  convergence to the sharp minima and thus tend to drive the model parameters to the flat region, similar to the goal of the adversarial training. Interestingly, unlike adversarial training methods that use a min-max optimization framework, our SharpCF  can be trained using standard Stochastic Gradient Descent, which significantly reduces the time complexity. Experimental results on real-world datasets show that our SharpCF outperforms the state-of-the-art adversarial collaborative filtering methods with almost no additional computational cost.

We summarize our contributions as follows:
\begin{itemize}[leftmargin=5mm]
    \item We have revisited the existing adversarial collaborative filtering methods and established their connection to recent Sharpness-aware Minimization. Through this, we have unveiled that adversarial training tends to favor flat minima over sharp ones, which results in better generalizability.
    \item We propose Sharpness-aware Collaborative Filtering (SharpCF) to introduce a novel trajectory loss that measures the alignment between current and past model states. Therefore, our SharpCF is able to avoid the need to solve a min-max optimization problem, and enables adversarial training without adding extra computational costs.
    \item Experimental results demonstrate that SharpCF outperforms the existing  collaborative filtering methods. Specifically, our SharpCF consistently outperforms the BPR  with an average improvement of 18.1\% and an average improvement of 6.82\%  over the APR. Additionally, in terms of time complexity, our SharpCF is comparable to the BPR and  has the same training speed as the BPR and is $2\times$ faster than the APR.
\end{itemize}

\section{Related Work}
In this section, we briefly review the related work on Collaborative Filtering and Adversarial Training. We also
highlight the differences between the existing efforts and our proposed method.
\subsection{Collaborative Filtering}

Collaborative filtering (CF) is a widely used technique in recommender systems, which plays an essential role in addressing the information overload problem for users~\cite{schafer2007collaborative}. The core idea of CF is that users tend to have similar preferences and hence, their opinions and behavior can be utilized to make recommendations in the near future.  One of the primary methods for CF is the  Matrix Factorization, which learns the latent user and item representations by factorizing the  observed interaction matrix~\cite{koren2009matrix,rendle2009bpr}. The predicted score of an unobserved user-item pair can be then derived by the similarity between the user and item representations, typically measured by the dot product.

Inspired by the success of deep neural networks, neural CF models have been proposed to learn more powerful user/item representations. For example,  The Wide\&Deep recommender model~\cite{cheng2016wide} combines linear models and deep neural networks to capture both memorization and generalization. He et al. \cite{he2017neural} propose a neural collaborative filtering model that replaces the dot product with a neural network architecture to model the user-item interactions. Another popular model is DeepFM~\cite{guo2017deepfm}, which combines factorization machines with neural networks to learn both low- and high-order feature interactions. XdeepFM~\cite{lian2018xdeepfm} extends DeepFM by introducing a novel cross network to capture more complex feature interactions. Recently, xLightFM~\cite{jiang2021xlightfm} greatly reduces the memory footprint of factorization machines via quantization techniques. In addition to factorization machines, graph neural networks  have also attracted increasing attention recently, and a number of graph-based CF models have been proposed, such as LightGCN~\cite{he2020lightgcn}.  These models have shown great success in recommendation tasks and have been applied in various domains~\cite{covington2016deep, koren2009matrix, rendle2009bpr,yeh2022embedding,wang2022improving,wang2023federated}.

However, noisy data (\textsl{e.g.}, false-positive feedback) can have a significant impact on the performance of recommender systems. Generally, noisy data can cause a bias towards popular items, which is very challenging for accurate and diverse recommendations~\cite{wang2021clicks,tian2022learning,chen2022denoising,wen2019leveraging,wang2021denoising}. To collect the data, prior studies consider incorporating more user feedback (\textsl{e.g.,} multi-type clicks, user textual comments) during training. Wen et al.~\cite{wen2019leveraging} suggest that training the recommender models uses
three kinds of scenarios: 
"click-complete", "click-skip", and "non-click"
ones, where last two kinds of items are both treated as negative samples. Julian et al.~\cite{mcauley2013hidden} combine the user latent factors with textual review to justify users' ratings. Nevertheless, additional feedback might be expensive or unavailable in many scenarios. Researchers have alternatively attempted to explore the adversarial training to mitigate the negative impact of noisy feedback   without additional
information~\cite{he2018adversarial,yuan2019adversarial,wu2021fight,chen2019adversarial,du2022understanding}.

\subsection{Adversarial Training}

Adversarial training has emerged as a promising technique to mitigate the negative effects of noisy data in recommender systems~\cite{he2018adversarial,yuan2019adversarial,wu2021fight,chen2019adversarial,du2022understanding,chen2022adversarial}. Adversarial training involves training the model on a combination of clean and adversarial data, where the adversarial data is generated by adding the worst-case perturbations to the input data~\cite{ian,he2018adversarial,chen2022adversarial}. The goal is to encourage the model to be robust to these perturbations, which in turn improves its ability to generalize to unseen data. For example, In APR~\cite{he2018adversarial}, adversarial training is used to perturb the embedding vectors of users and items in a Matrix Factorization model, resulting in a more robust and accurate recommendation system.  NMRN~\cite{wang2018neural} applies adversarial training to discover both long-term stable interests and short-term dynamic behaviors in   streaming recommender models.
ACAE~\cite{yuan2019adversarial}
combines collaborative auto-encoder with  adversarial training, which outperforms
highly competitive state-of-the-art recommendation methods. ATF~\cite{chen2019adversarial} reaps the benefits of adversarial training to improve the  context-aware recommendations. Typically, these methods employ a min-max optimization framework to alternatively learn adversarial perturbations and model parameters.  However, solving min-max frameworks incurs a two-fold computational overhead of the given base optimizer, making it not scalable to large datasets in practice.

Recently, several studies  have made attempts to develop the \textsl{fast} version of adversarial training~\cite{shafahi2019adversarial,andriushchenko2020understanding,Wong2020Fast}. One example is FreeAT~\cite{shafahi2019adversarial}, which removes the overhead cost of generating adversarial examples by reusing the gradient information computed during model training.  GradAlign~\cite{andriushchenko2020understanding} explicitly maximizes the gradient alignment inside the perturbation set. Despite their bleeding edge performance, those frameworks rely on a series of heuristics-based strategies, such as good initialization, large step size, and cyclic learning rate schedule. Recent studies indicate that these \textsl{free} strategies often lack of stability and  are prone to catastrophic overfitting issue~\cite{zhang2022revisiting}. Furthermore, the majority of fast adversarial training algorithms are intended for generating continuous adversarial examples (e.g., image pixels), rendering them unsuitable for information retrieval purposes due to their discrete space~\cite{he2018adversarial,wang2017irgan}.

In contrast, we propose a novel trajectory loss function that measures the alignment between current and past model states to reduce the complexity, which avoids solving min-max optimization. And we show that our trajectory loss function is  connected to the recent Sharpness-aware Minimization~\cite{foret2021sharpnessaware,kwon2021asam,andriushchenko2022towards}, which tends to seek flat minima, leading to better model generalization.

\section{PRELIMINARIES}
In this section, we first formulate the basic problem of collaborative filtering. Then, we briefly revisit the Bayesian Personalized Ranking (BPR)~\cite{rendle2009bpr} and  the Adversarial Personalized Ranking (APR)~\cite{he2018adversarial} for implicit recommendations.

\subsection{Collaborative Filtering}
In this paper, we focus  on the task of learning user preferences from implicit feedback.  Specifically,  the users' behavior data, e.g., click, view,  comment, purchase, etc., consists of  a set of users $\mathcal{U}=\{u\}$ and items $\mathcal{I}=\{i\}$, such that the set $\mathcal{I}_{u}^{+}$ represents the items that user $u$ has interacted with before,  whereas $\mathcal{I}_{u}^{-}=\mathcal{I}-\mathcal{I}_{u}^{+}$ represents unobserved items. In general, the unobserved interactions are not necessarily negative, but rather, the user may simply be unaware of them.  The goal of collaborative filtering is to estimate the user preference towards items.

The majority of CF methods learn each user and item into a low-dimensional latent space. Matrix Factorization   is widely acknowledged as the fundamental and most effective model in  recommendation. Specifically,  the representations of a user $u$ and an item $i$ can be obtained via embedding lookup tables:
	\begin{equation}
 \begin{aligned}
	\mathbf{e}_{u}=\text{lookup}(u), \qquad \mathbf{e}_{i}=\text{lookup}(i),
 \label{eq0}
 \end{aligned}
	\end{equation}
	where $u$ and $i$ denote the IDs of user and item; $\mathbf{e}_{u} \in \mathbb{R}^d$ and $\mathbf{e}_{i} \in \mathbb{R}^d$ are the embeddings of user $u$ and item $i$, respectively, and $d$ is the embedding size. These embeddings are intended to capture and retain the initial characteristics of both items and users, which can be updated during training.
Then, the predicted score is defined as the similarity between the user and item representations via dot product:
	\begin{equation}
		\hat{y}_{u i}={\mathbf{e}_{u}}^{T} \mathbf{e}_{i}.
  \label{eq2}
	\end{equation}

As for the learning objective, we next introduce the Bayesian Personalized Ranking (BPR) loss ~\cite{rendle2009bpr} and the  Adversarial Personalized Ranking (APR)~\cite{he2018adversarial} loss to train the model.

\subsection{Bayesian Personalized Ranking}

Bayesian Personalized Ranking (BPR)~\cite{rendle2009bpr} is a popular and effective pairwise method used in learning-to-rank   for recommender systems. Its primary goal is to optimize recommender models towards personalized ranking. BPR is particularly suitable for learning from implicit feedback, where the observed interactions are often incomplete and the unobserved ones are assumed to be ranked lower.

Unlike pointwise methods, which focus on optimizing each model prediction towards a predefined ground truth, BPR prioritizes maximizing the margin between an observed interaction and its unobserved counterparts. This margin-based approach allows BPR to perform well even when the number of negative samples is much larger than the number of positive samples. Formally, BPR aims to minimize the  following objective function:
	\begin{equation}
	\mathcal{L}_\text{BPR}( {\Theta})=-\sum_{(u, i, j) \in \mathbb{O}}\ln \sigma\left(\hat{y}_{u i}-\hat{y}_{u j}\right),
  \label{eq3}
	\end{equation}
	where $\mathbb{O}=\left\{(u, i, j) \mid u \in \mathcal{U} \wedge i \in I_{u}^{+} \wedge j \in  I_{u}^{-}\right\}$ denotes the pairwise
	training data, $\sigma(\cdot)$ is the sigmoid function, and ${\Theta}$ denotes model parameters. Typically, standard Stochastic Gradient Descent (SGD) is   used for its optimization. Once the parameters are obtained, a personalized ranking list for a user $u$ can be generated by evaluating the value of $\hat{y}_{ui}(\Theta)$ over all unobserved items $ i \in  I_{u}^{-}$.
 
However, training Matrix Factorization with the BPR loss is not robust as it is vulnerable to adversarial perturbations on the model parameters. Therefore, the model can easily learn a complex function, which may result in overfitting on the training data and poor generalization on unseen data.

\subsection{Adversarial Personalized Ranking}
To tackle above issue, Adversarial Personalized Ranking (APR)~\cite{he2018adversarial} intends to create an objective function that optimizes the recommender model for both personalized ranking and resistance to adversarial perturbations. To achieve this, APR injects adversarial perturbations ${\Delta}$  on  latent factors to quantify the loss of the model under perturbations on its parameters ${\Theta}$:
	\begin{equation}
	\begin{aligned}
	\hat{y}_{ui} (\Theta + \Delta) &= {(\mathbf{e}_{u} + {\Delta}_{u} )}^T (\mathbf{e}_{i}+{\Delta}_{i}),
	\end{aligned}
 \end{equation}
 where the  perturbation vectors $\Delta$ are coupled with their corresponding latent factors, \textsl{i.e.}, ${\Delta}_{u} \in \mathbb{R}^d$ denotes the perturbation vector for user latent vector $\mathbf{e}_u$. Adversarial perturbations aim to cause the largest influence on the model and are also known as worst-case perturbations.  Therefore, APR maximizes the BPR  loss to find the optimal adversarial perturbations:
\begin{equation}
	\begin{aligned}
\Delta_\text{adv} = &\argmax_\Delta  \mathcal{L}_\text{BPR} (\hat{\Theta} + \Delta),\\
 & s.t. \quad \|\Delta\|_2 \le \rho,
 \label{eq5}
	\end{aligned}
	\end{equation}
\noindent where $\rho$ controls the magnitude of adversarial perturbations, $\|\cdot\|_2$ denotes the $L_2$ norm, and $\hat{\Theta}$ is the intermediate model parameters. To this end, APR designs a new objective function that is both reasonable for personalized ranking and robust to adversarial perturbations. Formally, it minimizes  the adversarial BPR loss by jointly integrating Eq. (\ref{eq3}) and Eq. (\ref{eq5})  as follow~\cite{he2018adversarial,du2022understanding}:
	\begin{equation}
	\begin{aligned}
\mathcal{L}_\text{APR}(\Theta) = \mathcal{L}_\text{BPR} (\Theta) + \alpha \cdot \mathcal{L}_\text{BPR} ({\Theta} + \Delta_{adv}),\\
\text{where} \quad \Delta_{adv} = \argmax_{\|\Delta\|_2 \le \rho}  \mathcal{L}_\text{BPR} (\hat{\Theta} + \Delta),  
\label{eq6}
	\end{aligned}
	\end{equation}
\noindent where $\alpha$ controls the impact of the adversarial perturbations on the model optimization. In the extreme case where $\alpha = 0$, the APR  algorithm reduces to the original BPR  framework as defined in Eq.~(\ref{eq3}). Therefore, APR can be considered a generalization of BPR that takes into account the robustness of the models.

The above min-max objective function in Eq. (\ref{eq6}) can be expressed as playing a min-max game: the optimization of model parameters $\Theta$ serves as the minimizing player, while adversarial perturbations $\Delta$ act as the maximizing player. The two players alternate between playing this min-max game until convergence.  Nevertheless, APR still lacks theoretical guarantees of
why adding perturbations improves the model's generalizability. Additionally, solving min-max optimization is a time-consuming process. These shortcomings have motivated us to develop a new training method for personalized ranking. We next present our SharpCF that provides theoretical guarantees and requires nearly zero additional computational cost compared to the APR.

\section{The Proposed SharpCF}

In this section, we begin by simplifying the design of the APR loss function. Next, we establish a connection between the APR loss and recent research on Sharpness-aware Minimization~\cite{foret2021sharpnessaware,kwon2021asam,andriushchenko2022towards}. Building on this insight, we put forward Sharpness-aware Collaborative Filtering (SharpCF), which includes a novel trajectory loss that measures the alignment between the current and past model states. Importantly, we demonstrate theoretically that our trajectory loss performs a similar role as adversarial training on improving model generalization.

\subsection{Simplify APR}

Since the intermediate variable $\Delta$ maximizes the objective function that $\Theta$ minimizes, the optimization problem in Eq. (\ref{eq6}) can be  expressed as:
	\begin{equation}
	\begin{aligned}
\Theta^*, \Delta^* = \argmin_\Theta \max_{ \|\Delta\|_2 \le \rho}  \underbrace{\mathcal{L}_\text{BPR}(\Theta)}_{\text{the BPR loss}}  + \alpha \cdot \underbrace{ \mathcal{L}_\text{BPR} ({\Theta} + \Delta)}_{\text{the adversarial loss}}.
	\end{aligned}
	\end{equation}

One can adopt the classical gradient descent-ascent algorithm for min-max
optimization, which alternatively
updates one variable while fixing the other one. That is
\begin{equation}
	\begin{aligned}
\Delta^{(t+1)} &\gets \argmax_{\|\Delta\|_2 \le \rho}  \mathcal{L}_\text{BPR} ({\Theta}^{(t)} + \Delta), \\
\Theta^{(t+1)} &\gets \argmin_\Theta \mathcal{L}_\text{BPR}(\Theta) + \alpha \cdot  \mathcal{L}_\text{BPR} ({\Theta} + \Delta^{(t+1)}).
	\end{aligned}
	\end{equation}

Following~\cite{he2018adversarial}, we adopt the  Fast Gradient Sign Method to solve the inner maximization while apply the  standard Stochastic Gradient Descent to optimize the outer minimization. Clearly, we have the following observations: 1) While updating $\Delta$, only the adversarial loss contributes to the gradient $\nabla_\Delta \mathcal{L}_\text{BPR} ({\Theta}^{(t)} + \Delta)$; 2) While updating $\Theta$, the gradient of adversarial loss $ \nabla_\Theta \mathcal{L}_\text{BPR} ({\Theta} + \Delta^{(t+1)})$  already contains the gradient of BPR loss  $\nabla_\Theta \mathcal{L}_\text{BPR} ({\Theta})$ as the $\Delta^{(t+1)}$ is a constant.
Given this information, a natural question arises here: is it necessary to include the original BPR loss $\mathcal{L}_\text{BPR}(\Theta)$ in the min-max optimization?

To answer the above question, we conduct  a series of experiments using four public benchmark datasets: MovieLens1M, Gowalla, Yelp2018, and Amazon-Book (The detailed data description can be found in Sec 5). In the experiments, we fix the embedding size $d=128$, and the magnitude of adversarial perturbations $\rho=0.5$. Then we vary the regularization parameter $\alpha$   within the range of $\{0, 1e-2, 1e-1, 1, 1e1, 1e2, 1e3, +\infty\}$. We next compare the model training results using two different loss functions: Hybrid loss and Adversarial loss. The Hybrid loss is a combination of the BPR loss and the Adversarial loss with a range of $\alpha$ values except for $\{0,+\infty\}$, while the Adversarial loss completely eliminates the impact of the BPR loss by setting $\alpha=+\infty$. Our comparison of these two methods provided valuable insights into their respective impact for adversarial training. 

\begin{figure*}
	\begin{center}
		\includegraphics[width=15.0cm]{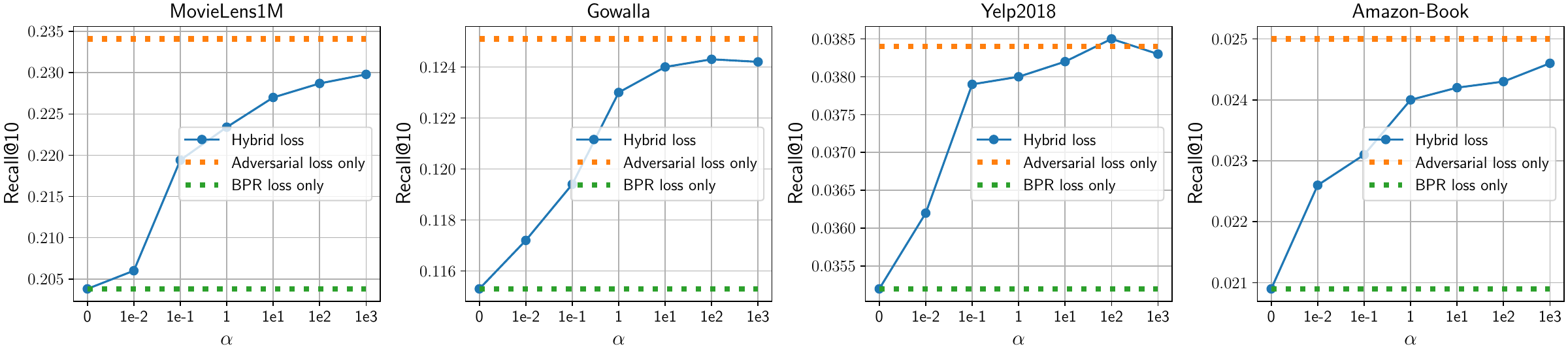}
	\end{center}
	\caption{The performance of APR with different values of $\alpha$. The Hybrid loss is a combination of the BPR loss and the Adversarial loss with a range of $\alpha$ values except for $+\infty$, while the Adversarial loss completely eliminates the impact of the BPR loss by setting $\alpha=+\infty$.}
	\label{layer}
\end{figure*}

As depicted in Figure \ref{layer}, choosing a non-zero value of $\alpha$ consistently outperforms the baseline that sets $\alpha=0$, \textsl{i.e.}, the model training only with the BPR loss. This means that the model gets benefits from the adversarial training.  Gradual improvements are commonly observed as $\alpha$ increases, particularly when $\alpha$ is smaller than $100$. However, future increases in $\alpha$ do not significantly improve or decrease the performance when $\alpha$ is larger than $100$. 

In the extreme case $\alpha=+\infty$, training with only the adversarial loss slightly improves performance, except for the Yelp dataset, but the difference between the Hybrid loss and the Adversarial loss is not statistically significant. Based on the above analysis, we can conclude that  the model benefits from a larger value of $\alpha$  and becomes rather insensitive when $\alpha$ is sufficiently large.  This implied that the Adversarial loss actually dominates the training, and the BPR loss can be safely removed without hurting the performance.

To this end, we can simplify the APR model in Eq. (\ref{eq6}) into the following:
	\begin{equation}
	\begin{aligned}
 \min_\Theta \max_{\|\Delta\|_2 \le \rho}   \mathcal{L}_\text{BPR} ({\Theta} + \Delta).
 \label{eq9}
	\end{aligned}
	\end{equation}

 We next connect the adversarial training to the   recent Sharpness-aware Minimization~\cite{foret2021sharpnessaware,kwon2021asam,andriushchenko2022towards}.

\subsection{Connect to Sharpness-aware Minimization}
Understanding the generalization of adversarial training  is critical as the  training
objective Eq. (\ref{eq9}) has multiple local optima that can perfectly fit the training data, but different 
optima lead to dramatically different generalization performance. To better understand the generalization of Eq. (\ref{eq9}), we decompose it as:
\begin{equation}
	\begin{aligned}
& \min_\Theta \mathcal{L}(\Theta) + \mathcal{R}(\Theta), \\
 \quad \text{where} \quad \mathcal{R}(\Theta) &= \max_{ \|\Delta\|_2 \le \rho}   \mathcal{L} 
 ({\Theta} + \Delta) - \mathcal{L}(\Theta),
 \label{eq10}
	\end{aligned}
	\end{equation}
where $\mathcal{L}$ is short for $\mathcal{L}_\text{BPR}$. Interestingly, we observe that the Eq. (\ref{eq10}) is the same as the Sharpness-aware Minimization~\cite{foret2021sharpnessaware}, where the term $\mathcal{R}(\Theta)$ captures the sharpness of $\mathcal{L}$ at $\Theta$ by measuring how quickly the training loss can be increased by moving $\Theta$ to a nearby region: $\{\Theta + \Delta| \Delta: \|\Delta\|_2 \le \rho \}$.

Therefore, rather than seeking model parameters $\Theta$ that have low training loss values in Eq. (\ref{eq3}), adversarial training in Eq. (\ref{eq10}) actually seeks out $\Theta$ whose entire neighborhoods have uniformly low training loss values. In other words,  adversarial training   favors \textsl{flat} minima rather than \textsl{sharp} minima, resulting in better generalizability.  One can further derive generalization bounds as:
	\begin{prop}
For any $\rho>0$, let $\mathcal{L}_\mathcal{D}$ be the expected loss and $\mathcal{L}_\mathcal{S}$ be the training loss, where  the training set $\mathcal{S}$ is drawn from data distribution  $\mathcal{D}$ with i.i.d condition, then 
\begin{equation}
		\label{prop1}
\mathcal{L}_\mathcal{D}(\Theta) \le \max_{  \|\Delta\|_2 \le \rho}   \mathcal{L}_\mathcal{S}  ({\Theta} + \Delta) + h(\|\Theta\|^2_2/\rho^2),
\end{equation}
where $h(\cdot): \mathbb{R}_+ \to \mathbb{R}_+$ is a strictly increasing function (\textsl{e.g.}, the weight decay $L_2$ norm).
	\end{prop}

The proof of the above inequality can be driven by using the PAC-Bayes theory~\cite{mcallester1999pac} (more details can be seen in Theorem 2 in ~\cite{foret2021sharpnessaware}). Above proposition provides a generalization upper bound to explain why the adversarial training can help improve generalization.  

However, both Adversarial Training~\cite{he2018adversarial} and  Sharpness-aware Minimization~\cite{foret2021sharpnessaware} adopt the   gradient descent-ascent framework, which requires two forward and backward passes on each sample in a batch, namely a namely a gradient ascent step to update  the perturbation $\Delta$ and a gradient descent step to update the current model $\Theta$. This doubling of computation time compared to the base optimizer makes them unsuitable for scaling to industry-scale datasets.

\subsection{Our SharpCF}

To overcome the computational bottleneck, we propose Sharpness-aware Collaborative Filtering (SharpCF), which includes a novel trajectory loss that measures the alignment between the current and past model states. We begin by approximate the inner maximization problem in Eq. (\ref{eq10}) via a first-order Taylor expansion of $\Delta$ as~\cite{foret2021sharpnessaware}:
\begin{equation}
	\begin{aligned}
\hat{\Delta} &= \arg \max_{ \|\Delta\|_2 \le \rho}   \mathcal{L} 
 ({\Theta} + \Delta) \\
 &\approx \arg \max_{\|\Delta\|_2 \le \rho}   \mathcal{L} 
 ({\Theta} ) + \Delta^T \nabla_\Theta  \mathcal{L} 
 ({\Theta} ) \\
 &= \arg \max_{\|\Delta\|_2 \le \rho}   \Delta^T \nabla_\Theta  \mathcal{L} ({\Theta} ) \\&= \rho \frac{\nabla_\Theta  \mathcal{L} 
 ({\Theta} )}{\|\nabla_\Theta  \mathcal{L} 
 ({\Theta} )\|_2}.
\end{aligned}
\end{equation}
With the approximated $\hat{\Delta}$,  for each batch size $\mathbb{B}$, we can rewrite its sharpness term $\mathcal{R}_{\mathbb{B}}(\Theta)$ as:
\begin{equation}
	\begin{aligned}
\mathcal{R}_\mathbb{B}(\Theta) &= \max_{ \|\Delta\|_2 \le \rho} \mathcal{L}_\mathbb{B}
 ({\Theta} + \Delta) - \mathcal{L}_\mathbb{B}(\Theta) \\
 &\approx \max_{\|\Delta\|_2 \le \rho} \mathcal{L}_\mathbb{B}
 ({\Theta} ) + \hat{\Delta}^T \nabla_\Theta  \mathcal{L}_\mathbb{B}(\Theta)  - \mathcal{L}_\mathbb{B}(\Theta) \\
 & = \rho \frac{\nabla_\Theta  \mathcal{L}_\mathbb{B} 
 ({\Theta} )^T}{\|\nabla_\Theta  \mathcal{L}_\mathbb{B} 
 ({\Theta} )\|_2}  \nabla_\Theta  \mathcal{L}_\mathbb{B}(\Theta) \\
 &=  \rho \|\nabla_\Theta  \mathcal{L}_\mathbb{B} 
 ({\Theta} )\|_2.
 \label{e12}
\end{aligned}
\end{equation}
This remarks that minimizing the sharpness term $\mathcal{R}_{\mathbb{B}}(\Theta)$ is equivalent to minimizing the $l_2$-norm of the gradient $\nabla_\Theta  \mathcal{L}_\mathbb{B}(\Theta)$, which is the same gradient used to minimize the vanilla loss $\mathcal{L}_\mathbb{B}(\Theta)$. However, directly optimizing the gradient norm involves second-order derivative information (\textsl{e.g.}, Hessian), which is computationally demanding. To overcome this challenge, we next
introduce a novel trajectory loss that measures the alignment between current and past model states~\cite{du2022sharpness}. Thus, the trajectory loss provides a way to optimize the gradient norm implicitly without requiring the computation of second-order derivatives. 

For current iteration $t$, we denote its pass trajectory of the model weights as $\mathbf{\Theta}=\{\Theta_1, \cdots, \Theta_{t-1}\}$, and $\Theta_t$ represents the current weights in the $t-$th iteration. Recall that standard SGD updates the weights as $\Theta_{t+1} = \Theta_t - \eta_t \nabla_{\Theta_t}  \mathcal{L}_{\mathbb{B}_t} 
 ({\Theta}_t ) $. For current batch $\mathbb{B}_t$ and model state $\Theta_t$, as $\mathcal{R}_{\mathbb{B}_t}(\Theta_t)$ is always non-negative, and thus we have:
 \begin{equation}
	\begin{aligned}
\arg \min_{\Theta_t} \quad \mathcal{R}_{\mathbb{B}_t}(\Theta_t) &\Leftrightarrow \arg \min_{\Theta_t} \quad  \eta_t\cos(\Phi_t)\mathcal{R}_{\mathbb{B}_t}(\Theta_t) \mathcal{R}_{\mathbb{B}_t}(\Theta_t)\\
&\Leftrightarrow \arg \min_{\Theta_t} \quad  \eta_t\cos(\Phi_t)\mathcal{R}_{\mathbb{B}_t}(\Theta_t) \mathcal{R}_{\mathbb{B}_t}(\Theta_t) \\
&+ \sum_{i<t}  \eta_i\cos(\Phi_i)\mathcal{R}_{\mathbb{B}_t}(\Theta_i) \mathcal{R}_{\mathbb{B}_i}(\Theta_i)
\\
& = \sum_{i=1}^t  \arg \min_{\Theta_t} \quad   \eta_i \cos(\Phi_i)\mathcal{R}_{\mathbb{B}_t}(\Theta_i) \mathcal{R}_{\mathbb{B}_i}(\Theta_i)\\
& = \underset{\Theta_i \sim \mathrm{Unif}( \mathbf{\Theta})}
{\mathop{\mathbb{E}}}[ \eta_i\cos(\Phi_i)\mathcal{R}_{\mathbb{B}_t}(\Theta_i) \mathcal{R}_{\mathbb{B}_i}(\Theta_i)],
\label{e13}
\end{aligned}
\end{equation}
where $\Theta_i \sim \mathrm{Unif}( \mathbf{\Theta})$ denotes that $\Theta_i$ is uniformly distributed in the set $\mathbf{\Theta}$, and $\mathbb{E}[\cdot]$ denotes the expectation; $\Phi_i$ is the angle between the gradients that are computed using the mini-batches $\mathbb{B}_i$ and $\mathbb{B}_t$ and $\cos(\Phi_i)=1$. Note that $ \eta_i\cos(\Phi_i)\mathcal{R}_{\mathbb{B}_t}(\Theta_i) \mathcal{R}_{\mathbb{B}_i}(\Theta_i)$ becomes a constant with respect to variable $\Theta_t$ for all $i \neq t$. By substituting Eq. (\ref{e12}) into Eq (\ref{e13}), we have following:
 \begin{equation}
	\begin{aligned}
\arg& \min_{\Theta_t}  \quad \mathcal{R}_{\mathbb{B}_t}(\Theta_t) \Leftrightarrow  \underset{\Theta_i \sim \mathrm{Unif}( \mathbf{\Theta})}
{\mathop{\mathbb{E}}}[ \eta_i\cos(\Phi_i)\mathcal{R}_{\mathbb{B}_t}(\Theta_i) \mathcal{R}_{\mathbb{B}_i}(\Theta_i)] \\
& =  \underset{\Theta_i \sim \mathrm{Unif}( \mathbf{\Theta})}
{\mathop{\mathbb{E}}}[\rho^2 \eta_i\cos(\Phi_i) \|\nabla_{\Theta_i}  \mathcal{L}_{\mathbb{B}_t} 
 ({\Theta_i} )\|_2  \|\nabla_{\Theta_i}  \mathcal{L}_{\mathbb{B}_i} 
 ({\Theta}_i )\|_2] \\
 & = \underset{\Theta_i \sim \mathrm{Unif}( \mathbf{\Theta})}
{\mathop{\mathbb{E}}}[\rho^2 \eta_i\nabla_{\Theta_i}  \mathcal{L}_{\mathbb{B}_t} 
 ({\Theta_i} )^T   \nabla_{\Theta_i}  \mathcal{L}_{\mathbb{B}_i} 
 ({\Theta}_i )] \\
 & \approx  \rho^2 \underset{\Theta_i \sim \mathrm{Unif}( \mathbf{\Theta})}
{\mathop{\mathbb{E}}}[  \mathcal{L}_{\mathbb{B}_t} 
 ({\Theta_i} )-  \mathcal{L}_{\mathbb{B}_t} 
 ({\Theta}_{i+1} )] \\
 & = \frac{\rho^2}{t-1} [ \mathcal{L}_{\mathbb{B}_t} 
 ({\Theta_1} )-  \mathcal{L}_{\mathbb{B}_t} 
 ({\Theta}_2 ) + \cdots + \mathcal{L}_{\mathbb{B}_t} 
 ({\Theta_{t-1}} )-  \mathcal{L}_{\mathbb{B}_t} 
 ({\Theta}_t )] \\
 & = \frac{\rho^2}{t-1} [ \mathcal{L}_{\mathbb{B}_t} 
 ({\Theta_1} )-   \mathcal{L}_{\mathbb{B}_t} 
 ({\Theta}_t )].
\end{aligned}
\end{equation}
 We remark that minimizing the sharpness term $\mathcal{R}_{\mathbb{B}_t}(\Theta_t)$ is equivalent to minimizing the loss difference $ \mathcal{L}_{\mathbb{B}_t} 
 ({\Theta_1} )-   \mathcal{L}_{\mathbb{B}_t} 
 ({\Theta}_t )$. To avoid the second term $-   \frac{\rho^2}{t-1} \mathcal{L}_{\mathbb{B}_t} 
 ({\Theta}_t )$ cancel out a partial of the vanilla loss, we replace $\mathcal{L}_{\mathbb{B}_t} 
 ({\Theta_1} )-   \mathcal{L}_{\mathbb{B}_t} 
 ({\Theta}_t )$ with its  $l_2$ norm. In addition, as the model parameters $\Theta_1$ contains less information, for current epoch  $e$, we only track the loss trajectory in the past $E$ epochs: $\{\Theta_{e-E}, \cdots, \Theta_{e}\}$.
 
 To this end, the mini-batch loss of our SharpCF at $e$-th epoch is defined as follow:
	\begin{equation}
	\mathcal{L}_\text{SharpCF}( {\Theta_e})= \underbrace{  \mathcal{L}_\mathbb{B} ( {\Theta_e})}_{\text{the BPR loss}} + \frac{\lambda}{|\mathbb{B}|}  \underbrace{   \| \mathcal{L}_{\mathbb{B}} 
 ({\Theta_e} )-   \mathcal{L}_{\mathbb{B}} 
 ({\Theta}_{e-E} )\|^2_2}_{\text{the trajectory loss}},
 \label{eq16}
	\end{equation}
 where $\lambda$ is a regularized parameter that is used to   balance the vanilla loss with the trajectory loss. Meanwhile, $\mathcal{L}_{\mathbb{B}} 
 ({\Theta}_{e-E} )$ represents the loss of the batch $\mathbb{B}$ in $E$ epochs ago, which is being recorded during training.

 Nonetheless, both  $\mathcal{L}_{\mathbb{B}} 
 ({\Theta}_{e-E} )$ and  $\mathcal{L}_{\mathbb{B}}({\Theta}_{e} )$  involve the utilization of the negative sampling technique to generate the negative samples as described in Eq. (\ref{eq3}). In practice, the same batch of positive user-item pairs can be utilized for $\mathcal{L}_{\mathbb{B}} 
 ({\Theta}_{e-E} )$ and  $\mathcal{L}_{\mathbb{B}} 
 ({\Theta}_{e} )$ through an indexing method, but distinct negative pairs are generated  due to the random negative sampling approach. To address this issue, we initially train the model with BPR, and subsequently monitor the trajectory after a predefined epoch $E_\text{start}$.

After the pretrained epoch $E_\text{start}$, the user/item representations become more reliable and stable, and the BPR loss enforces a large margin between positive pairs and negative pairs, \textsl{i.e.}, $\hat{y}_{ui} \gg \hat{y}_{uj}$, for $i \in I_{u}^{+} \wedge j \in I_{u}^{-}$. As such, we  empirically find that it is sufficient to only track the predicted scores of positive pairs in the mini-batch $\hat{\mathbf{Y}}_{\mathbb{B}} 
 ({\Theta}_{e-E} )$ and  $\hat{\mathbf{Y}}_{\mathbb{B}} 
 ({\Theta}_{e} )$  as described in Eq. (\ref{eq2}). By doing so, our  empirical loss function of Eq. (\ref{eq16}), which is simple yet effective, can be expressed as follows:
\begin{equation}
	\mathcal{L}^{\text{emp}}_\text{SharpCF}( {\Theta_e})= {  \mathcal{L}_\mathbb{B} ( {\Theta_e})} + \frac{\lambda}{|\mathbb{B}|}  {   \| \hat{\mathbf{Y}}_{\mathbb{B}} 
 ({\Theta_e} )-   \hat{\mathbf{Y}}_{\mathbb{B}} 
 ({\Theta}_{e-E} )\|^2_2},
 \label{eq17}
	\end{equation}
 
 Essentially, our loss is applied to slow down the rate of change of the training loss to prevent convergence to sharp local minima.  Algorithm \ref{alg1} summarizes the overall training of SharpCF.

	\begin{algorithm}
		\DontPrintSemicolon % Some LaTeX compilers require you to use \dontprintsemicolon instead
		\KwIn{ The training data $\mathbb{O}$, the regularizer $\lambda$, the number of epoch $\overline{E}$, the started trajectory loss epoch $E_\text{start}$, the epoch window $E$.}
		
		Initialize model parameters ${\Theta}$;\;
		\For{ $e \gets 1$ \textbf{to} $\overline{E}$}{
			\For{\text{each mini-batch} $\mathbb{B} \subset \mathbb{O} $}{
             Compute the predicted scores for positive pairs  $\hat{\mathbf{Y}}_\mathbb{B} ( {\Theta_e})$;\;
              Compute the predicted scores for negative pairs  $\hat{\mathbf{N}}_\mathbb{B} ( {\Theta_e})$ ;\;
				Compute the BPR loss  $\mathcal{L}_\mathbb{B} ( {\Theta_e})$ based on $\hat{\mathbf{Y}}_\mathbb{B} ( {\Theta_e})$ and $\hat{\mathbf{N}}_\mathbb{B} ( {\Theta_e})$;\;
              Cache the loss $\hat{\mathbf{Y}}_{\mathbb{B}} 
 ({\Theta}_{e-E} )$ in $E$ epochs ago for same batch $\mathbb{B}$;\;
               \eIf{$e > E_\text{start}$}{
$\mathcal{L}^{\text{emp}}_\text{SharpCF}( {\Theta_e})= \mathcal{L}_\mathbb{B} ( {\Theta_e})  + \frac{\lambda}{|\mathbb{B}|}     \| \hat{\mathbf{Y}}_{\mathbb{B}} 
 ({\Theta_e} )-  \hat{\mathbf{Y}}_{\mathbb{B}} 
 ({\Theta}_{e-E} )\|^2_2$;\;
               }{
             $\mathcal{L}^{\text{emp}}_\text{SharpCF}( {\Theta_e})= \mathcal{L}_\mathbb{B} ( {\Theta_e})$;
               }
               Update the model weights $\Theta_e$ by using SGD;\;
			}
		}
		\caption{SharpCF}
  \label{alg1}
		\KwOut{ The optimal model parameters $\Theta^*$.}
	\end{algorithm}

\subsubsection*{\textbf{Model Complexity}}
For time complexity, Unlike adversarial training~\cite{he2018adversarial,foret2021sharpnessaware} that  requires two forward and backward passes to alternatively update the  auxiliary variable $\Delta$ and model parameters $\Theta$, Our SharpCF can  update  $\Theta$ by using standard SGD. This makes the time complexity of SharpCF similar to that of the original BPR model~\cite{rendle2009bpr}, with almost no additional computational cost required to record the loss trajectory.

In terms of memory complexity, APR~\cite{he2018adversarial} requires $\mathcal{O}((|\mathcal{U}| + |\mathcal{I}|)d)$, where $|\mathcal{U}|$ and $|\mathcal{U}|$ represent the number of users and items, respectively, and $d$ is the embedding size. In contrast, our SharpCF needs extra memory to cache the loss trajectory $\mathcal{O}(E|\mathbb{O}|)$, where $E$ is the window size of the trajectory and  $|\mathbb{O}|$ is the number of training data.  However, since the window size of the trajectory $E$ is typically a small constant like $3$ or $5$, the memory complexity of SharpCF is negligible compared to BPR~\cite{rendle2009bpr}. Overall, our proposed SharpCF method offers a computationally efficient and memory-friendly alternative for collaborative filtering tasks.

\section{EXPERIMENTS}
In this section, we present the results of our extensive experiments on four public datasets, aimed at validating the effectiveness of SharpCF. Firstly, we describe our experimental settings, and then we compare the overall top-$K$ recommendation performance of SharpCF with other CF methods. Next, we showcase the loss landscape of our SharpCF, which reveals that the adversarial training effectively smooths the loss landscape to achieve flat minima. Finally, we investigate the performance of sparse recommendations to verify the generalizability of different CF methods.

	\subsection{Experimental Settings}
	\subsubsection{Datasets} We use four public benchmark datasets for evaluating recommendation performance:
	\begin{itemize}
		\item \textbf{Movielens-1M}\footnote{https://grouplens.org/datasets/movielens/} is a widely used dataset for recommendations. This dataset  contains $1,000,209$ anonymous ratings of approximately $3,900$ movies 
made by $6,040$   users who joined in $2000$.
		\item \textbf{Gowalla}\footnote{https://github.com/kuandeng/LightGCN/tree/master/Data} is a location-based social networking website where users share their locations by checking-in. It mainly collects the check-ins of these users over the period from Feb. 2009 to Oct. 2010.
		\item \textbf{Yelp2018}\footnote{https://www.yelp.com/dataset} is released by the {Yelp} challenge that consists of a subset of the businesses, reviews, and user data. The {Yelp2018} version is used in the experiments.
		\item \textbf{Amazon-Book}\footnote{https://jmcauley.ucsd.edu/data/amazon/}  comprises a vast corpus of user reviews, ratings, timestamps, and product metadata gathered from Amazon.com. For our experiments, we select the largest category available, namely Book.
	\end{itemize}
For the MovieLens1M dataset, we consider all ratings as implicit feedback, where each rating score is converted to either 1 or 0 to indicate whether a user rated a movie. For sparser datasets such as Gowalla, Yelp2018, and Amazon-Book, we use the 10-core setting   to ensure that all users and items have at least 10 interactions~\cite{he2020lightgcn}. Table 1 provides a summary of the dataset statistics.
	\begin{table}[]
		\caption{Dataset statistics.}
		\label{my-label}
		\begin{tabular}{ccccc}
		\toprule[1.2pt]
			Dataset & \#User & \#Items & \#Interactions & Density \\ \hline
			MovieLens1M    & 6,040    & 3,900     & 1,000,209      & 4.190\%\\      
			Gowalla    & 29,858    & 40,981     & 1, 027, 370      & 0.084\%\\      
			Yelp2018   & 31, 668    & 38, 048     & 1, 561, 406      & 0.130\%\\      
			Amazon-Book   & 52, 643    & 91, 599    & 2, 984, 108     & 0.062\%\\      \toprule[1.2pt]
		\end{tabular}
	\end{table}

	\subsubsection{Baselines} As our primary goal of this work is to give a deep insight about the adversarial training and accelerate its computational overhead,  we mainly compare with the two popular baselines:
	
	\begin{itemize}
		\item \textbf{BPR}~\cite{rendle2009bpr}: A classic model that seeks to optimize the Bayesian personalized ranking loss. We employ Matrix Factorization as its preference predictor.
		\item \textbf{APR}~\cite{he2018adversarial}: Similar to BPR, it also chooses the Matrix Factorization model and considers the adversarial perturbations during training.
	\end{itemize} 

For BPR, APR, and SharpCF, we  choose the basic Matrix Factorization as their backbones due to its simplicity and effectiveness. However, it is worth mentioning that SharpCF, similar to BPR,  is compatible with any encoders, such as Graph Neural Networks~\cite{he2020lightgcn,chen2023sharp}. We leave this extension for further work since our focus is not on developing a new encoder for recommendation in this study. In contrast, we pay more attention to explaining and accelerating the adversarial collaborative filtering.

	\subsubsection{Implementation Details} We implement our SharpCF model in PyTorch on NVIDIA Tesla V100 32GB machines. For all models, the embedding dimension $d$ of users and items (\textsl{e.g.}, in Eq. (\ref{eq0})) is set to 128.  We initialize the hyper-parameters for APR as suggested in the original paper and then fine-tune them to optimize performance.  For our SharpCF, we choose an epoch window $E=3$. As suggested by APR~\cite{he2018adversarial}, we set $E_\text{start}= 200$, which means we first train the model with BPR only for the first 200 epochs to warm up, and then continue training with APR or SharpCF in the experiments.

We adopt two popular top-$K$ metrics, Recall and Normalized Discounted Cumulative Gain (NDCG)~\cite{he2020lightgcn}, for evaluation purposes. The default value of $K$ is set to $[10,20]$, and we report the average of Recall$@10$ and NDCG$@10$ over all users in the test set. During inference, we consider items that the user has never interacted with in the training set as candidate items. All models predict the users' preference scores over these candidate items and rank them based on the computed scores to calculate Recall$@10$ and NDCG$@10$. We independently repeated the experiments five times and report the averaged results.

\subsection{Experimental Results}

\begin{table*}[]
\caption{The performance of different baselines in terms of Recall@K, NDCG@K, and the duration of each epoch in seconds. RI means Relative Improvement w.r.t. baselines.}
\label{tab:my-table}
\scalebox{1.1}{\begin{tabular}{ccccccc}
\hline
Dataset                       & Metrics   & BPR    & APR    & SharpCF         & RI w.r.t. BPR & RI w.r.t. APR \\ \hline
\multirow{5}{*}{Movielens-1M} & Recall@10 & 0.2038 & 0.2273 & \textbf{0.2354} & +15.5\% & +3.56\% \\
                              & NDCG@10   & 0.3396 & 0.3705 & \textbf{0.4123} & +21.4\% & +11.3\% \\
                              & Recall@20 & 0.3025 & 0.3126 & \textbf{0.3179} & +5.09\% & +1.70\% \\
                              & NDCG@20   & 0.3387 & 0.3566 & \textbf{0.3949} & +16.6\% & +10.7\% \\
                              & Time per epoch   & 16.5s & 30.5s & 17.1s & - & - \\\hline
\multirow{5}{*}{Gowalla}      & Recall@10 & 0.1153 & 0.1242 & \textbf{0.1296} & +12.4\% & +4.35\% \\
                              & NDCG@10   & 0.1258 & 0.1375 & \textbf{0.1456} & +15.7\% & +5.89\% \\
                              & Recall@20 & 0.1674 & 0.1784 & \textbf{0.1838} & +9.80\% & 3.03\% \\
                              & NDCG@20   & 0.1404 & 0.1524 & \textbf{0.1594} & +13.5\% & +4.59\% \\ 
                              & Time per epoch   & 28.3s & 50.1s & 28.7s & - & - \\ \hline
\multirow{5}{*}{Yelp2018}     & Recall@10 & 0.0352 & 0.0397 & \textbf{0.0411} & +16.8\% & +3.53\% \\
                              & NDCG@10   & 0.0446 & 0.0477 & \textbf{0.0526} & +17.9\% & +10.3\% \\
                              & Recall@20 & 0.0591 & 0.0673 & \textbf{0.0686} & +16.1\% & +1.93\% \\
                              & NDCG@20   & 0.0519 & 0.0571 & \textbf{0.0608} & +17.1\% & +6.48\% \\
                              & Time per epoch  & 45.8s & 88.6s & 46.1s & - & - \\ \hline
\multirow{5}{*}{Amazon-Book}  & Recall@10 & 0.0207 & 0.0241 & \textbf{0.0267} & +29.0\% & +10.8\% \\
                              & NDCG@10   & 0.0223 & 0.0259 & \textbf{0.0294} & +31.9\% & +13.5\% \\
                              & Recall@20 & 0.0366 & 0.0421 & \textbf{0.0451} & +23.2\% & +7.13\% \\
                              & NDCG@20   & 0.0286 & 0.0329 & \textbf{0.0363} & +26.9\% & +10.3\% \\ 
                              & Time per epoch   & 127.8s & 230.1s & 128.9s & - & -\\ \hline
\end{tabular}}
\end{table*}

\subsubsection{Overall Performance}

In this study, we compare our proposed method SharpCF with BPR and APR on four real-world datasets. Table~\ref{tab:my-table} summarizes the experimental results, including the running time per epoch. Based on our experiments, we have made the following two observations:

\begin{itemize}[leftmargin=5mm]
\item 
Although the underlying recommender model remains the same (e.g., Matrix Factorization), both APR and SharpCF outperform BPR by leveraging adversarial training. APR achieves this by explicitly introducing worst-case perturbations, while our SharpCF implicitly smooths the learning trajectory to achieve the same effect. Both methods have the potential to smooth the loss landscape to reach out the flat minima, leading to better performance. These findings suggest that the way of training a recommender model is a critical factor in the recommendation process, and one can easily modify the vanilla loss function to enhance performance further.

\item Our proposed SharpCF has shown improvements in all comparisons. It consistently outperforms the BPR by up to ~30\%, with an average improvement of 18.1\% and a standard deviation of 7\% across all datasets. While the improvements over the  APR are not as significant as those over the BPR, our SharpCF still shows positive results, with an average improvement of 6.82\% and a standard deviation of 3.81\% over the APR. Presumably, this is because our SharpCF does not require solving the min-max minimization,  which avoids the risk of   getting stuck at saddle points that often exhibit large variances, as explained in the original APR paper.
\end{itemize}

\begin{figure*}
	\begin{center}
		\includegraphics[width=15.0cm]{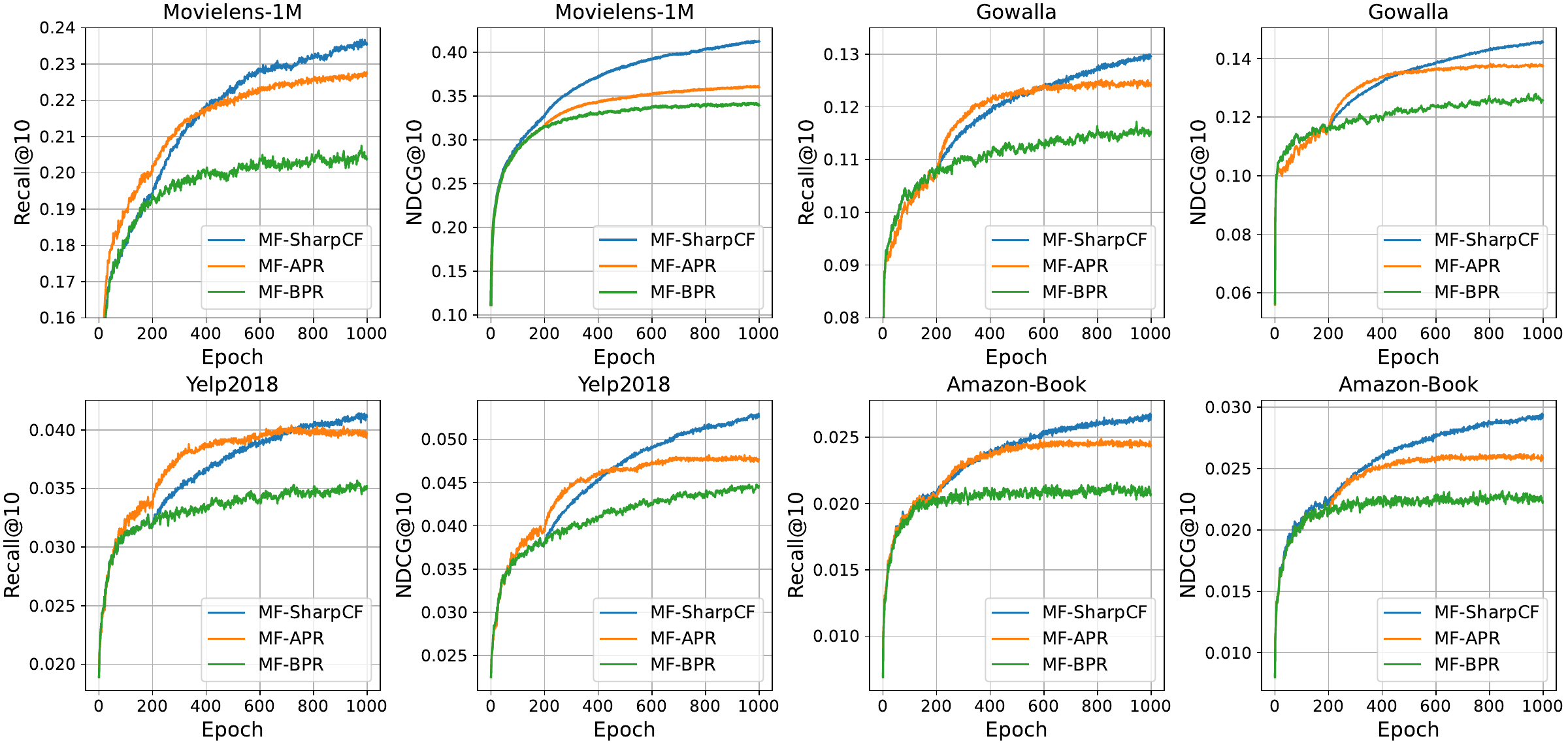}
	\end{center}
	\caption{Training curves of BPR, APR, and SharpCF on different datasets.}
	\label{curve}
\end{figure*}

Figure~\ref{curve}  depicts the training curves of various baselines for four datasets. After pretraining for 200 epochs, we observe that continued training of APR and SharpCF results in significant improvement, whereas further training of BPR leads to only marginal gains. For instance, in the case of Amazon-Book, BPR achieves a maximum Recall@10 score of around $0.0207$, which is then boosted to $0.0267$ by training with SharpCF, leading to a relative improvement of approximately $29\%$. Table~\ref{tab:my-table} also shows running  time elapsed for training per epoch of BPR, APR and SharpCF. In general, the training time for BPR and SharpCF is almost equal, while the computational time for APR is roughly twice as expensive as these two approaches. For example, for the largest dataset Amazon-Book, the training times per epoch for BPR, APR, and SharpCF are around $127.8$s, $230.1$s, and $128.9$s, respectively.

Overall, the experimental results demonstrate the superiority of our proposed SharpCF. Specifically, it outperforms the BPR and APR across four datasets, while maintaining comparable complexity to the BPR model. These appealing properties make our SharpCF practical for industry-scale applications.

\subsubsection{Loss Landscapes}
Prior works have demonstrated a strong correlation between the flatness of the loss landscape and the generalizability and robustness of a model. In Section 4.2, we establish a connection between Adversarial Training and Sharpness-aware Minimization to explain their generalizability.   To verify this assumption, we visualize the loss landscapes of the baselines and our proposed method on four different datasets. Following the methodology outlined in~\cite{li2018visualizing}, give a well-train model $\Theta$, we compute
the loss values when moving the model parameters $\Theta$ along a random direction $t \in [-2,2]$ to generate a 1D loss landscape. 

\begin{figure*}
	\begin{center}
		\includegraphics[width=15.0cm]{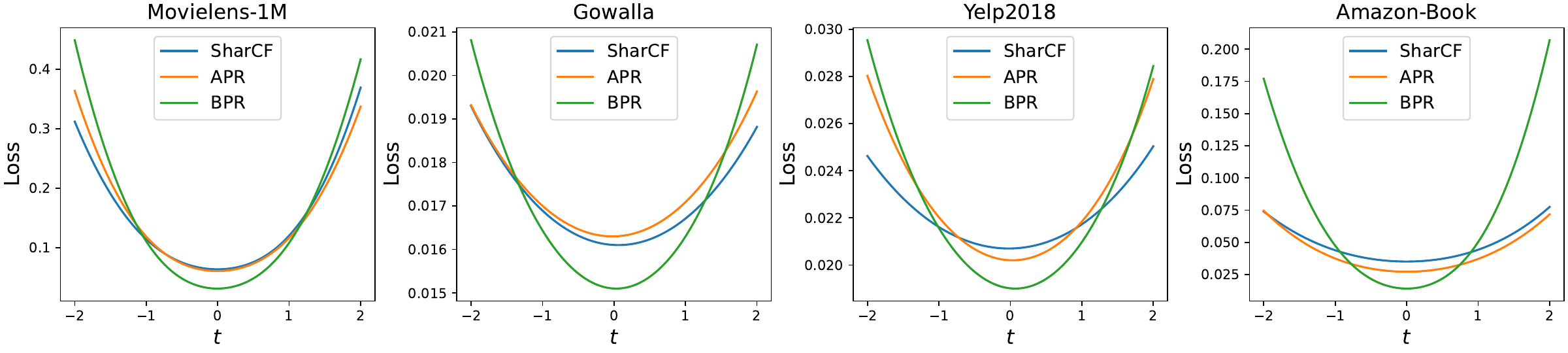}
	\end{center}
	\caption{Training curves of BPR, APR, and SharpCF on different datasets.}
	\label{fig:loss_landscape}
\end{figure*}

From Figure~\ref{fig:loss_landscape}, we observe that BPR has a sharper local minima in the loss landscape as compared to APR and SharpCF across all datasets. In other words, the adversarial training techniques, including APR and SharpCF, prefer to generate flatter loss landscapes.
A flatter weight loss surface often leads to a smaller gap between training and testing performances, thereby improving the robustness of the model's generalization capabilities~\cite{li2018visualizing,foret2021sharpnessaware,kwon2021asam,andriushchenko2022towards}. Rather than interpreting adversarial training as a min-max game aimed at enhancing robustness and generalizability, as  done in APR, we provide an alternative perspective by showing that adversarial training tends to achieve a flatter weight loss landscape, thereby reducing the robust generalization gap.

\subsection{Further Probe}
In this section, we conducted a series of detailed analyses on the proposed SharpCF to confirm its effectiveness. We report the results only on the Gowalla and Amazon-Book datasets, as the observations are similar on the other two datasets and thus omitted here.
\subsubsection{Long-tail Recommendation}

\begin{figure}
	\begin{center}
		\includegraphics[width=7.5cm]{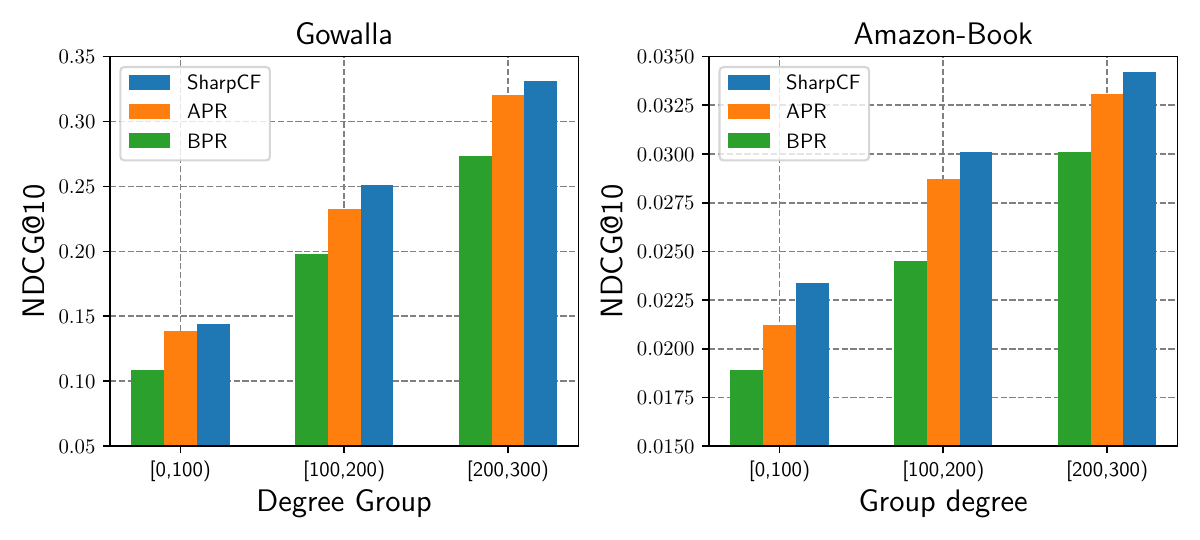}
	\end{center}
	\caption{The performance of different models with respect to item degree on Gowalla and Amazon-Book.}
	\label{sp}
\end{figure}
To further investigate the model's generalization, we grouped items by degree and visualized the average performance of each group. We mainly focused on Gowalla and Amazon-Book since they are sparser than the other two datasets. Additionally, we explored the item degree within the ranges of [0,100), [100, 200), and [200, 300) to focus on long-tail items. Figure~\ref{sp} displays the performance of different models with respect to item degrees. As shown, SharpCF and APR achieved higher performance for low-degree items. This indicates that both APR and SharpCF are capable of providing high-quality recommendations even with sparse interaction data, thanks to smoothing the loss landscape and better generalization.

\subsubsection{The Impact of the Coefficient $\lambda$}
\begin{figure}
	\begin{center}
		\includegraphics[width=6.6cm]{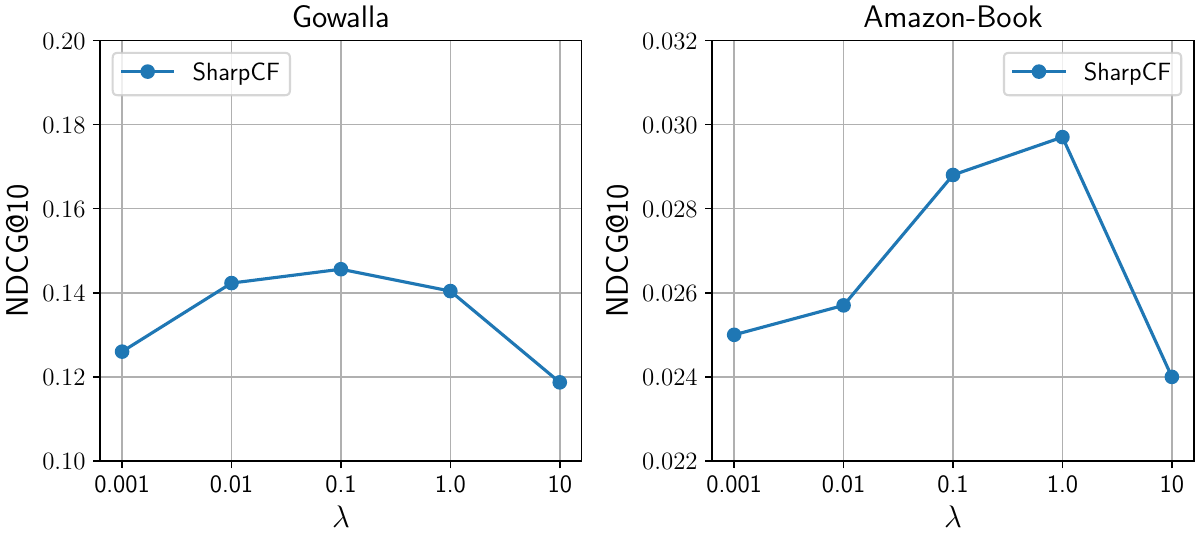}
	\end{center}
	\caption{The performance of SharpCF for different settings of $\lambda$.}
	\label{lamb}
\end{figure}

In the objective function of SharpCF defined in Eq. (\ref{eq17}), the coefficient $\lambda$ is used to balance the two losses: the original task loss and the trajectory loss. To analyze the influence of $\lambda$, we vary its value in the range of $0.001$ to $10$ and reported the experimental results in Figure \ref{lamb}. The results indicate that an appropriate value of $\lambda$ can effectively improve the performance of SharpCF. Specifically, when the hyper-parameter $\lambda$ is set to around $0.1$ or $1.0$, the performance becomes better on both datasets, revealing that tracking the learning trajectory is valuable for improving performance. However, when $\lambda$ is set to around $10.0$, the performance of our SharpCF drops quickly, suggesting that too strong regularization on the trajectory loss will negatively affect normal training of the model and is not encouraged in practice.

\subsubsection{The Impact of the Epoch Window Size $E$}
\begin{figure}
	\begin{center}
		\includegraphics[width=6.6cm]{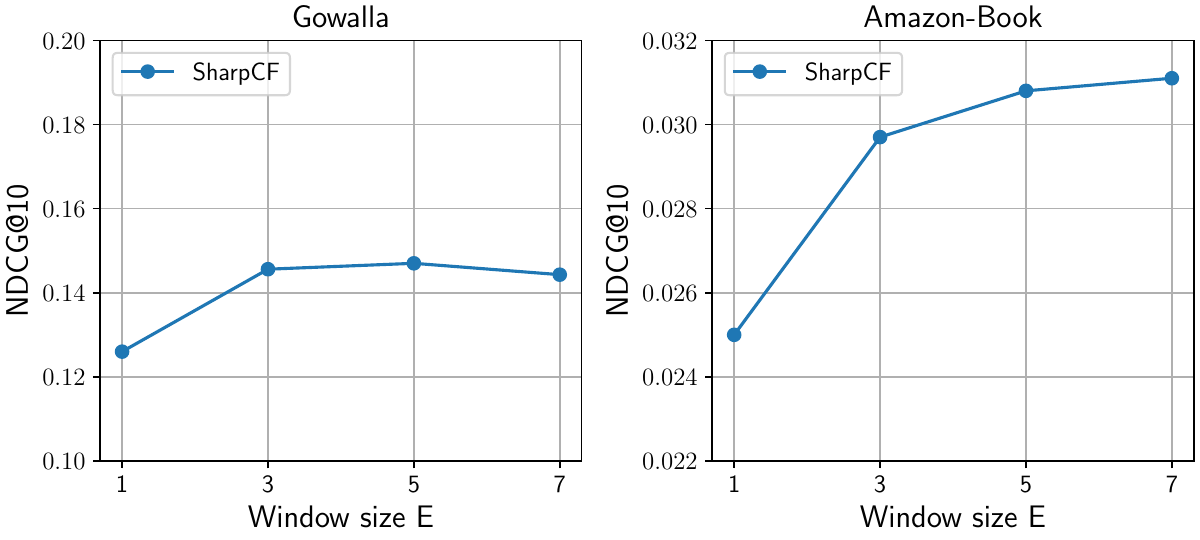}
	\end{center}
	\caption{The performance of SharpCF for different settings of the epoch window size $E$.}
	\label{e}
\end{figure}

In our proposed SharpCF model, the epoch window size $E$ in Eq. (\ref{eq17}) is a crucial hyper-parameter that affects its performance. A larger value of $E$ means that the SharpCF will track the long-range trajectory to smooth the loss landscape. To examine the impact of $E$ on SharpCF, we vary it from 1 to 7 and evaluate the model's performance on different datasets. From Figure \ref{e}, the results show that the performance of SharpCF is relatively stable across different settings of $E$. In some cases, increasing the value of $E$ can gradually boost the performance on the Amazon-Book dataset.  It should be noted that tracking longer trajectories in SharpCF requires more memory to cache the model states. While this is a cheaper alternative to the APR method, it still needs to be considered in real-world applications. In our experiments, we found that choosing an epoch window size of $E=3$ provides good recommendation performance with the reasonable memory requirement.

\section{Conclusion}

In this paper, we first revisit the existing adversarial collaborative filtering methods and demonstrate  that adversarial training favors flat minima over sharp ones, which results in better generalizability. We then propose our Sharpness-aware Collaborative Filtering (SharpCF), a simple yet effective method that conducts adversarial training without extra computational cost over the base optimizer. Our proposed method has shown superior performance on four public real-world datasets compared to current adversarial collaborative filtering methods with reasonable time complexity.

In our future work, we aim to expand our assessment of the effectiveness of our adversarial training approach in diverse recommendation tasks, including fairness recommendation. Additionally, we plan to extend the capabilities of our framework to establish it as a versatile training technique that improves model generalization.

\bibliographystyle{ACM-Reference-Format}
\bibliography{reference}

\end{document}